\documentstyle[hip-artc]{article}  
\volnumber{9}  \edyear{2004}  \frompage{000} \topage{000}                
\recrevdate{March 2004}                                              
\def\rightmark{} 
\def\leftmark{Alexandru Jipa et al.}
 
\title{Overview of the results from the BRAHMS experiment}
\authors{
{\twerm Alexandru Jipa for the BRAHMS Collaboration$^1$  %
}\\[2.812mm]
{\normalsize
\hspace*{-8pt}$^1$ Faculty of Physics, University of Bucharest, \\
P.O.Box MG-11, RO 077125 Bucuresti-Magurele, ROMANIA\\[0.2ex]
}}

\abstract{An overview of the most important experimental results obtained in the first 
three running years with the BRAHMS experiment from Brookhaven National Laboratory (USA) is presented. 
The design of the experiment permits to measure the interesting physical quantities 
in a large ranges of rapidity and transverse momentum. Therefore, properties of hadron production vs rapidity and
transverse momenta are presente.}
\PACS{$^b$}
 
\begin{document}
 
\maketitle
 
\section{Introduction}
The study of the ultra-relativistic heavy ion collisions has as one 
of its goals the recreation, in laboratory, of the conditions after Big Bang, at a few $\mu$s. 

According to the present cosmological theories, at this moment 
in the early Universe, a deconfined state of quarks and gluons has been existed. To obtain such 
phase of the nuclear matter, in laboratory, a transition to the hadron to quarks and gluons is 
necessary. Therefore, high nuclear temperatures and densities are requested. The nucleus-nucleus 
collisions at relativistic and ultra-relativistic energies can offer such conditions.

BRAHMS experiment was placed at two o'clock on the ring of the Relativistic Heavy Ion 
Collider from Brookhaven National Laboratory (USA). This experiment was designed to gather 
information on the interesting physical quantities characterising various emitted particles in 
ultrarelativistic nuclear collisions as functions of transverse momentum, $p_{T}$, and rapidity, y. 

The yields as function of rapidity offer information on the nuclear density into a given collision, 
as well as on the produced entropy. The information on the collision dynamics and the thermalisation 
degree is obtained from the spectral shapes of the interesting physical quantities and their dependencies 
on rapidity. Important information on the collision dynamics is obtained from collision centrality 
measurements, and the information from the times in the reaction is included in the high pT parts of the
spectra.

In the first three years of running different experimental data have been collected.  In the present paper 
a few of the most important experimental results obtained in Au-Au collisions at $\sqrt{s_{NN}}=130GeV$ and $\sqrt{s_{NN}}=200GeV$, 
as well as in d-Au collisions at $\sqrt{s_{NN}}=200GeV$ are included. The experimental data obtained in p-p 
collisions at $\sqrt{s_{NN}}=200GeV$ are used as reference data.

The experimental results include information on charged particle multiplicity, pseudorapidity and 
rapidity distributions, participants, energy density, nuclear temperatures and radial flow, antiparticle 
to particle ratios, Coulomb momentum, chemical potentials, entropy per baryon and high $p_{T}$ suppression. 
All quantities offer information on the behaviour of the highly excited and dense nuclear matter formed 
in the overlapping region of the two colliding nuclei, as well as on the possible formation of some new 
forms of the nuclear matter. Some comments and comparisons with different assumptions are included, too.

Before the presentation of the experimental results, a short description of the BRAHMS experiment is given.

\section{The BRAHMS experiment}  
The BRAHMS ({\bf B} road {\bf RA} nge {\bf H} adron {\bf M} agnetic {\bf S} pectrometers) Experiment is one of the experiments  
at Relativistic Heavy Ion Collider (RHIC) from Brookhaven National Laboratory (BNL) 
\cite{bib1,bib2,bib3,bib4,bib5}.

In BRAHMS experiment two different regimes can be studied, namely: (i) a baryon poor region with 
a high energy density, created at mid-rapidity; (ii) a region at high 
rapidities, near the initial nuclei, very rich in baryons at relatively high temperature. The BRAHMS experiment well
identified charged hadrons - $\pi^{+}$, $\pi^{-}$, $K^{+}$, $K^{-}$, $p$ and $\bar {p}$  - over a wide range of rapidity and 
transverse momenta at several energies and beams available at RHIC can be measured.  

Taking into account the accelerator system structure and the size of the experimental halls the 
BRAHMS spectrometers are small solid angle devices. They provide semi-inclusive measurements in very 
different experimental conditions.

\begin{figure}
\vspace*{-0.65cm}
                 \insertplot{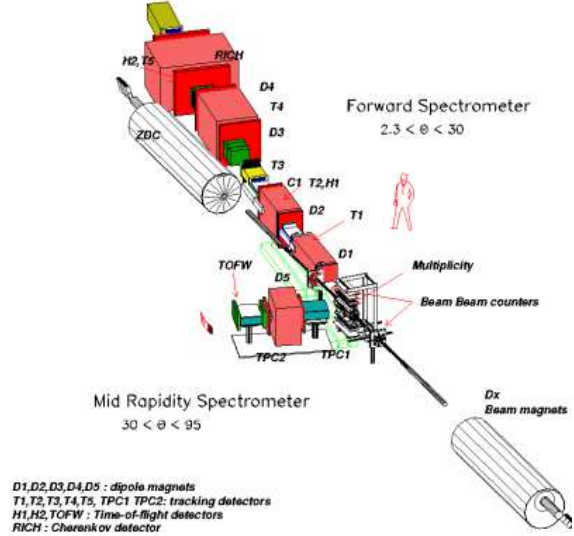}
\vspace*{-0.6cm}
\caption[]{A top view of the BRAHMS Experiment}
\label{fig1}
\end{figure}

The experiment was designed with two moveable magnetic spectrometers, namely: 
(a) Forward Spectrometer (FS); (b) Mid Rapidity Spectrometer (MRS) \cite{bib1,bib2,bib3,bib4,bib5}.
The Forward Spectrometer covers the angular region $2.3^{0}<\theta<30^{0}$, and the Mid Rapidity Spectrometer covers the angular 
region $30^{0}<\theta<95^{0}$. 
The pseudorapidity ranges covered by the two spectrometers are: $1.3<\eta<4.0$, $0.1<\eta<1.3$, respectively. To the two magnetic 
spectrometers three event characterisation detector systems have been added; they provide global information. The three
detectors systems are the following: Beam-Beam Counters (BBC), multiplicity detector and Zero Degree 
Calorimeters (ZDC). Together, the three detector systems will provide centrality coverage in the mid 
rapidity region, namely: $-2.2<\eta<2.2$, in the region $3.2<\mid{\eta}\mid<4.3$, as well as at $0^{0}$. A top view of the BRRAHMS experiment 
is shown in Fig.1.

\section{Particle identification and momentum determination}
Particle identification can be done using different ways. First of all, the detectors from the two spectrometers can be 
used to combine the time-of-flight measurements in the two hodoscopes (H1 and H2 in Fig.1) with measurements in the threshold 
Cherenkov counter (C1) and the Ring Imaging Cherenkov Counter (RICH). The 40 scintillator slats in the first hodoscope and the 
32 scintillator slats in the second hodoscope - instrumented with photomultiplier tubes (PMT) at each end - allow the particle 
identification in the momentum range 1-20 GeV/c, in Forward Spectrometer. The time-of-flight array in Forward Spectrometer is 
place at 8.6 m from the nominal vertex position. Mid Rapidity Spectrometer uses for particle identification two time projection 
chambers (TPC1 and TPC2), a dipole magnet (D5) and a time-of-flight wall (TOFW) with 83 scintillator slats, placed at 4.3 m from 
the nominal vertex position. In the two arms of the BRAHMS experiment the overall time resolution is around 90 ps. 
A second time-of-flight wall was introduced in 2003. In the Forward Spectrometer the separation of kaons and protons is acheived 
in the momentum range $p<4.5$ GeV/c, and in Mid Rapidity Spectrometer in the momentum range $p<2.2$ GeV/c. The time-of-flight 
performances for MRS and for FS are presented in Fig.2.

\begin{figure}
\vspace*{-0.65cm}
                 \insertplot{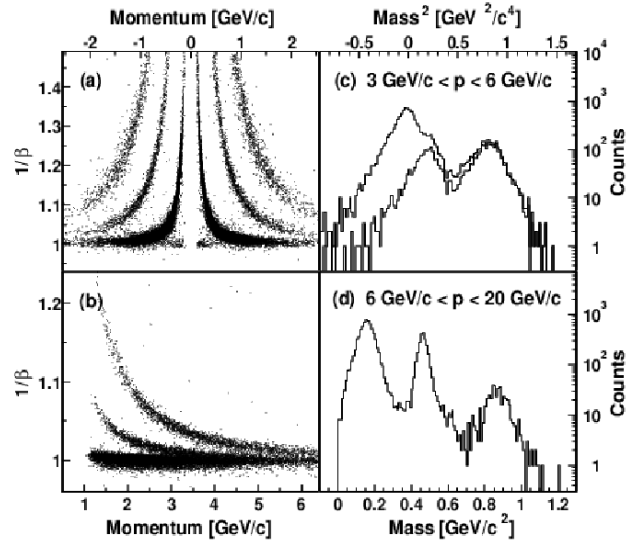}
\vspace*{-0.6cm}
\caption[]{Time-of-flight performance for MRS at 90 degrees (a) and for FS at 12 degrees (b). Mass-squared spectrum in the FS (c). Mass spectrum in FS using the RICH}
\label{fig2}
\end{figure}

Another way is related to the momentum spectra and effective mass. For fixed rapidities the $m^{2}$ spectra for positive and negative 
particles have been obtained. The following relation has been used:

\begin{equation}
m^{2}=p^{2}(\frac{t^2}{L^2}-1)
\end{equation}

where p is the particle momentum, t is the time-of-flight and L is the flight distance.

Particle momenta are determined by projecting the straight-line tracks as reconstructed in the two TPCs to the magnet and calculating 
the bending angle at matched tracks using an effective edge approximation. For obtaining momentum spectra the vertex dependent 
acceptance maps have been used. Reactions with a vertex in the range from -15 cm to +15 cm were used in the analysis. The momentum 
resolution obtained with the BRAHMS experiment $\frac{{\delta}p}{p}\leq 0.02$ 
\cite{bib2,bib3,bib4,bib5}. 

\begin{figure}
\vspace*{-0.6cm}
                 \insertplot{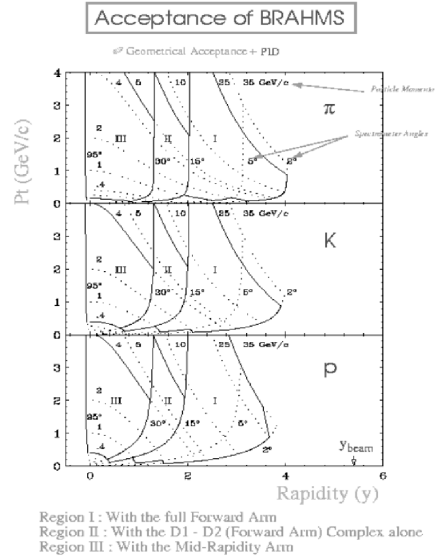}
\vspace*{-0.9cm}
\caption[]{Acceptance maps for BRAHMS experiment}
\label{fig3}
\end{figure}

The acceptances of the two spectrometers for different identified particles is shown in Fig.3. It is important to stress here 
that the acceptances for the two spectrometers, at a given field, for positive charged particles, are equal to the acceptances for 
negative charged particles at the opposite polarity of the given field. Therefore particle ratios can be obtained.

\section{The pseudorapidity and rapidity distributions and charged particle multiplicity}

For obtaining global information the Beam-Beam Counters (BBC), multiplicity detector and Zero Degree Calorimeters (ZDC) have been used. 
The Beam-Beam Counters provide the initial trigger and vertex information. They provide the start time for the time-of-flight 
measurements, too. Each of the elements of the BBCs has 3 cm and 4 cm UVT Cherenkov radiator read out by PMTs. The right side of 
the BBCs consists from 36 elements and the left side consists from 44 elements; they are positioned at $\pm2.2m$ from the nominal vertex 
interaction. The multiplicity detector provides information on the collision centrality. 24 segmented Si-detectors with 168 channels 
(SiMA = Si Multiplicity Array) and 40 scintillator squared tiles with 12 cm length (TMA = Tiles Multiplicity Array) are included in 
this hybrid detector. The pseudorapidity range covered by the multiplicity detector is $-2.0\leq\eta\leq2.0$. Common devices for all experiments at RHIC 
are the Zero Degree Calorimeters \cite{bib6}. At the BRAHMS experiment they are placed at $\pm14m$ from the nominal vertex interaction, 
situated behind the two DX beam-line magnets. For Au-Au collisions the two ZDCs offer information on the collision centrality 
measuring the forward going spectator neutrons. The coverage provides by the global detectors is in the region $-2.2\leq\eta\leq2.2$, 
in the region $3.2\leq\mid\eta\mid4.3$, as well as at $0^0$. They have been used in the three runs to measure the overall charged particle 
production and to determine collision centrality. SiMA and TMA can be hit by many different particles, including background particles. 
Therefore, to obtain the multiplicity information it is necessary to convert the deposited energy into a number of hits. The energy loss signal is transformed to charged 
particle multiplicity dividing total deposited energy to the expected deposited average energy by one particle in a tile. Conversion 
factors are evaluated from GEANT simulations. The background contributions are considered. They are between $10\%$ for SiMA elements 
and $30\%$ for TMA elements. All detector elements were calibrated in energy using pulsers, cosmic rays and sources measurements.

The pseudorapidity distributions, $\frac{dN}{d\eta}$, has been obtained from the measured energy depositions corrected using 
the above given corrections (Fig.4a). Selecting appropriate 
ranges in the multiplicity spectrum (Fig.4b), different cuts in collision centrality have been performed \cite{bib7,bib8}. Usually, the 
cuts in collision centrality are expressed in terms of the fraction of the nuclear reaction cross section by normalization to 
the integral of the TMA spectrum obtained, with a near minimum bias trigger. This trigger requires energy deposition in each of the 
two ZDCs above 25 GeV. An additional condition is that at least one tile having a hit. Almost $97\%$ of the nuclear interaction cross 
section can be selected by this requirement.

\begin{figure}
\vspace*{-.45cm}
\begin{center}
\begin{tabular}{cc}

\epsfxsize=6.2cm\epsffile{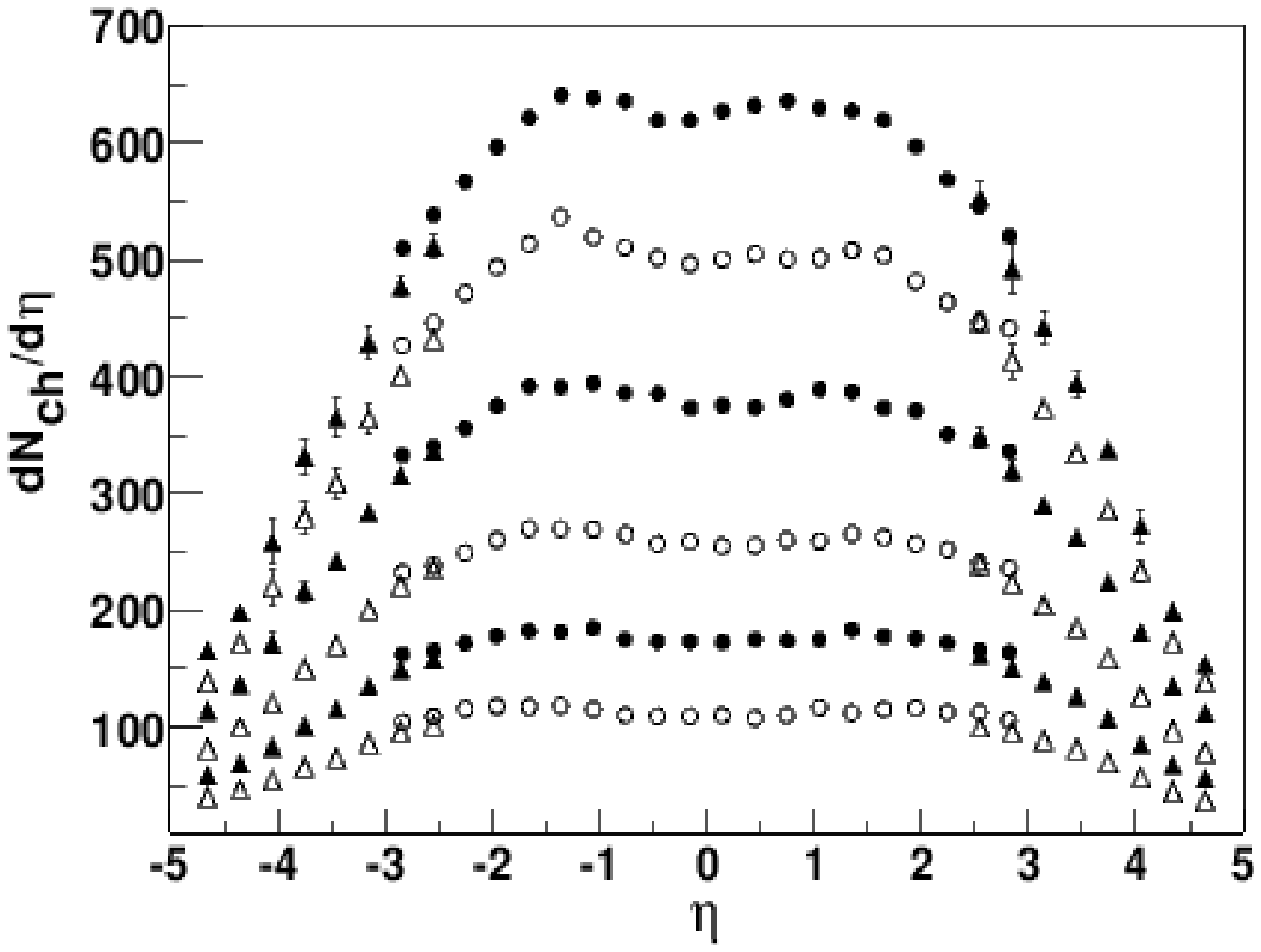}
&
\epsfxsize=6.2cm\epsffile{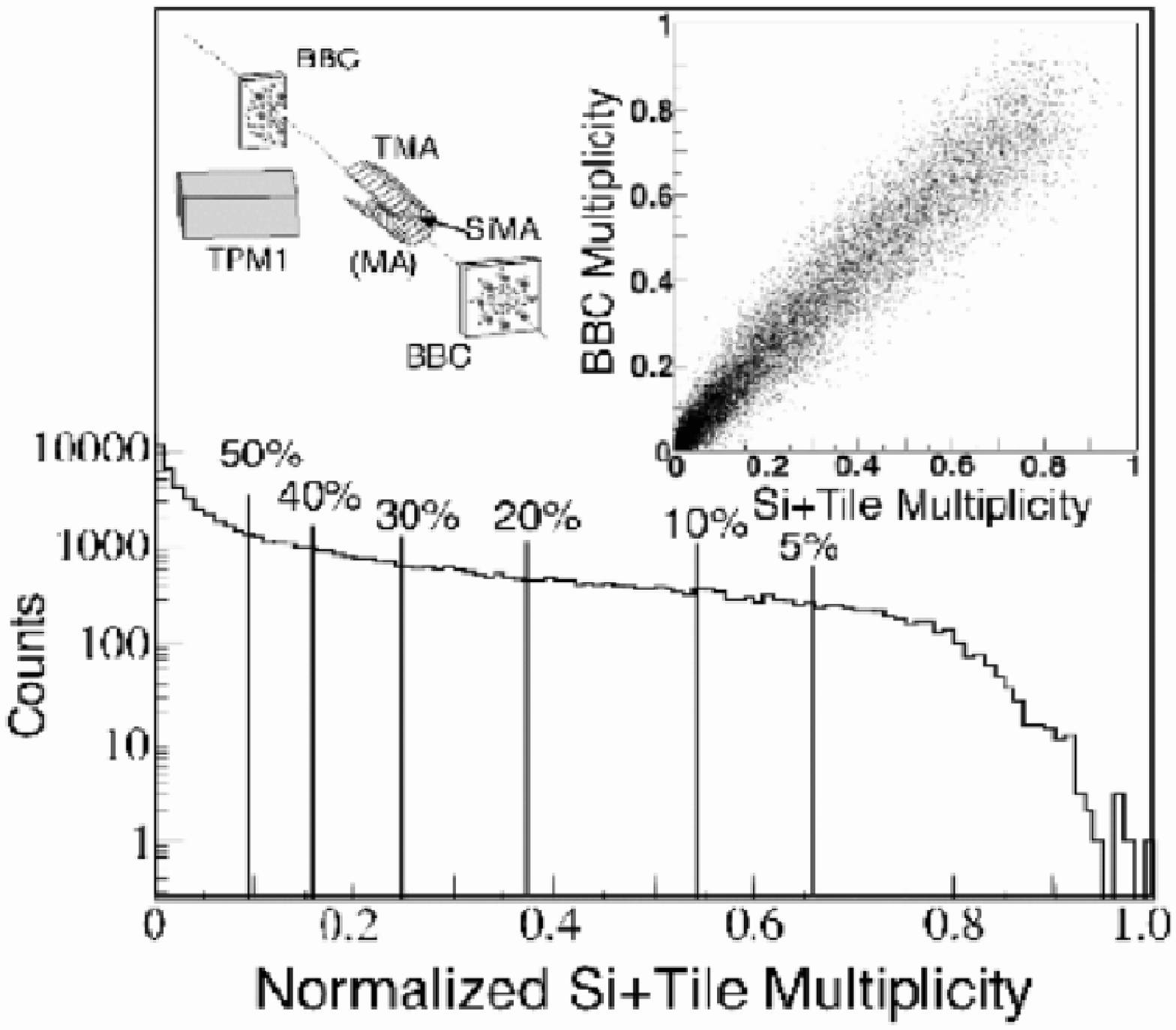} \\[0.1ex]
\end{tabular}
\vspace{-0.9cm}
\end{center}
\caption[]{(a) Pseudorapidity distribution of charged particles in Au-Au collisions at 200 AGeV. (b) Multiplicity spectrum for Au-Au collisions at 130 AGeV}
\label{fig4}
\end{figure}

In Au-Au collisions at $\sqrt{s_{NN}}=130$ GeV the charged particle multiplicity on the whole pseudorapidity range was 
$N^{130}_{ch}=4050\pm300$, and for the same collisions, at $\sqrt{s_{NN}}=200$ GeV the charged particle multiplicity was 
$N^{200}_{ch}=5100\pm300$.

For the most central Au-Au collisions (0-5\%), the values obtained from pseudorapidity distributions, at $\eta=0.0$, are: 
$\frac{dN}{d\eta}|^{130}_{ch}=550\pm30$, $\frac{dN}{d\eta}|^{200}_{ch}=610\pm50$, respectively. 
These values are in good agreement with HIJING code predictions \cite{bib10,bib9,bib4}. The full widths at the half maximum are:
$\Delta\eta^{130}_{ch}=7.6\pm0.7$, $\Delta\eta^{200}_{ch}=7.9\pm1.0$, respectively.

\begin{figure}
\vspace*{-0.6cm}
                 \insertplot{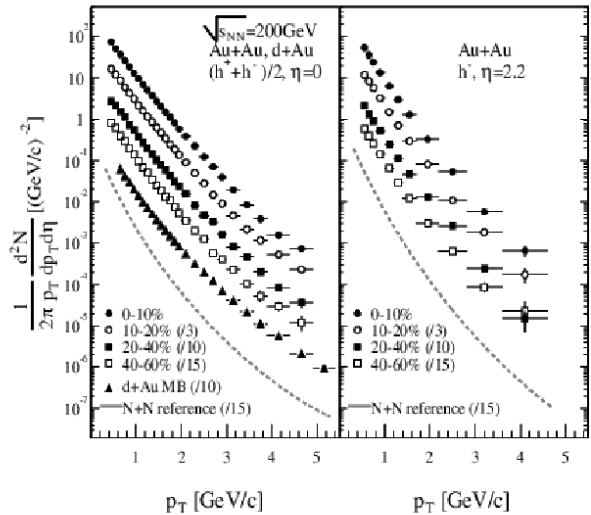}
\vspace*{-0.9cm}
\caption[]{Invariant spectra of charged hadrons from Au-Au collisions at 200 AGeV for 2 pseudorapidities: (a) 0.0, (b) 2.2, respectively. The results for d-Au collisions are included, too. }
\label{fig5}
\end{figure}

It is important to stress that similar values are obtained in the full pseudorapidity range. This fact
indicates the existence of a large plateau for particle production in very central collisions (6\%). Also, this value for $\eta=0.0$ is 
in good agreement with the value reported by PHOBOS Collaboration, value obtained in similar conditions \cite{bib11}.

The rapidity or pseudorapidity distribution can be used to estimate energy densities at the meson production, according 
Bjorken's formulae \cite{bib12}:

\begin{equation}
\epsilon=\frac{3}{2}\frac{<p_{T}>}{\tau{\pi}R^{2}}\frac{dN}{d\eta}
\end{equation}

For Au-Au collisions at $\sqrt{s_{NN}}=130$ and $\sqrt{s_{NN}}=200$ the estimated energy densities are $3.6 GeV/Fm^3$ and $4.5 GeV/Fm^3$, 
respectively.

\begin{figure}
\vspace*{-0.6cm}
                 \insertplot{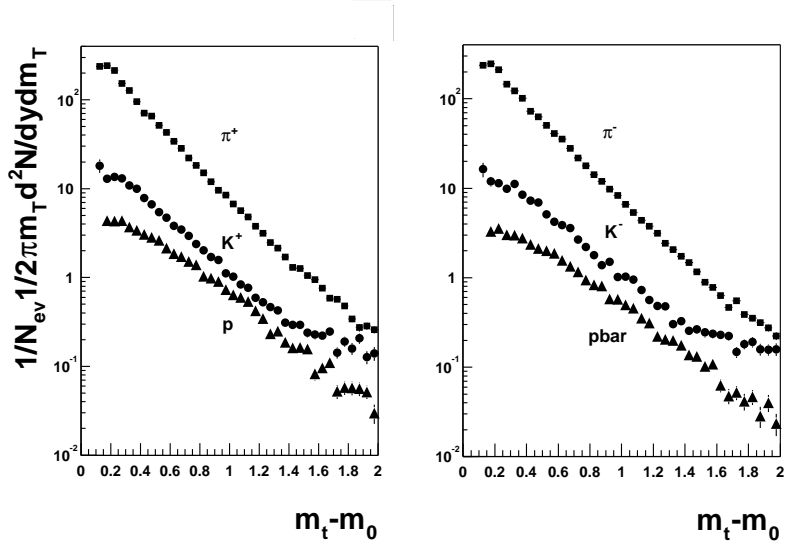}
\vspace*{-0.5cm}
\caption[]{Transverse mass spectra for charged pions, kaons, protons and antiprotons for the most central Au-Au collisions at 200AGeV, at y=0}
\label{fig6}
\end{figure}

Taking into account that the pseudorapidity distribution resembles Bjorken boost invariant assumption hydrodynamic 
calculations can be considered further. Therefore, some considerations on momentum spectra and the inverse slope are included.

\section{Momentum spectra and some dynamic information}
Usually, the high momentum particles are considered as a good probe to investigate the conditions prevailing early in 
the evolution of a system. Such particles can be associated with different production mechanisms like jet production; they
can loss energy due to induced gluon radiation passing through a medium with a high density of color charges. As a 
consequence, a depletion of the high transverse momentum component can by observed in the spectra.

In this paper the positive identified hadron momentum spectra, for 2 pseudorapiditiies ($\eta=0.0$, $\eta=2.2$, respectively), as well 
as those of their antihadrons are included (Fig.5). For each peseudorapidity different collision centralities are 
considered. These momentum spectra have been obtained as was presented previously. The invariant inclusive spectra of 
negative hadrons and positive hadrons, respectively, at $\eta=0.0$, permitted the estimation of the average values of the transverse 
momentum, for negative and positive hadrons. They are the following: $<p_{T}>_{h^{-}}=0.480GeV/c$, $<p_{T}>_{h^{+}}=0.450GeV/c$, respectively. 
These values are higher than those obtained by extrapolation at other energies \cite{bib13,bib15}; for example, at ISR and UA1 energies 
average transverse momentum is $<p_{T}>_{ex}=0.365GeV/c$.

\subsection{Inverse slope versus mass and centrality}

The transverse momentum and transverse mass spectra have been used to estimate the inverse slope parameters or nuclear 
temperatures, T, other physical quantity in the study of the conditions in early in the evolution of a system.

The transverse mass spectra for the hadrons detected with the BRAHMS experiment - for the most 
central Au-Au collisions at $\sqrt{s_{NN}}=200$GeV(0-10\%), at y=0 - are presented in Fig.6. For fitting the invariant transverse mass spectrum 
the following relation has been used:
\begin{equation}
\frac{1}{N_{ev}}\frac{1}{2{\pi}m_{T}}\frac{dN}{dm_{T}dy}=\frac{1}{2\pi}\frac{dN}{dy}\frac{1}{T(T+m_{0})}e^-{\frac{m_{T}-m_{0}}{T}}
\end{equation}
Here, $\frac{dN}{dy}$ is considered as a free parameter, for different centrality cuts.
The experimental results for the slope parameter T, for 4 different centrality cuts are included in Table I. The 
observed dependence of the temperature on the particle mass (T increases with the increase of the hadron mass), as 
well as the its dependence on the collision centrality (T increase with the increase of the collision centrality) 
is an indication of the radial expansion. Experimental values of the temperatures are higher than the values 
obtained from the simulations with HIJING and UrQMD codes \cite{bib10,bib14}.

\begin{table}
\vspace*{-12pt}
\caption[]{Experimental values of the temperatures for Au-Au collisions at 200AGeV and y = 0.0}
\vspace*{-14pt}
\begin{center}
\begin{tabular}{llllll}
\hline\\[-10pt]
Centrality & 0-10\% & 10-20\% & 20-40\% & 40-60\% \\ 
\hline\\[-10pt]
$T(\pi^{+})$ & 243.4$\pm$0.9 &  245.5$\pm$1.1 & 241.2$\pm$1.2 &  230.3$\pm$2.0 \\
$T(\pi^{-})$ & 242.6$\pm$0.9 &  241.6$\pm$1.1 & 240.7$\pm$1.2 &  231.5$\pm$1.9 \\
$T(K^{+})$ & 303.6$\pm$3.3 &  302.6$\pm$4.1 & 292.5$\pm$4.5 &  281.3$\pm$10.2\\
$T(K^{-})$ & 306.1$\pm$3.4 &  310.6$\pm$4.2 & 300.6$\pm$4.7 &  288.5$\pm$9.0 \\
$T(p)$ & 360.4$\pm$3.3 & 353.6$\pm$3.6 & 344.1$\pm$4.5 & 317.4$\pm$9.8 \\
$T(\bar{p})$ & 358.9$\pm$3.6 & 342.0$\pm$3.8 & 335.1$\pm$4.8 & 306.2$\pm$11.4\\
\hline 
\end{tabular}
\end{center}
\end{table}

From Table I it can observe that the pion's temperature is significantly higher than its rest mass. 

\subsection{Collective transverse flow}

The mass dependence of the slope parameters provides evidence of the collective flow. The following relation between
the slope parameter and particle mass has been proposed \cite{bib15}:

\begin{equation}
T=T_{fo}+m<v_{T}^{2}>
\end{equation}

where $T_{fo}$ is the freeze-out temperature, considered as the temperature for that the particles cease to interact with 
each other, and $<v_{T}>$ is the average collective flow velocity. The freeze-out temperature is related to the thermal, random 
processes, and the velocity is related to the collective processes. 

A fit with this formulae leads at freeze-out temperatures around 220 MeV, for all collision centralities, and collective 
flow velocities around 0.37c. Therefore, different relationships among these physical quantities have been proposed in the 
frame of different models \cite{bib16,bib17}. 

The experimental results for freeze-out temperature and transverse collective flow velocity, obtained in the framework of 
a blast wave model, are included in Table II. A very slow dependence of the freeze-out temperature on the collision centrality 
is observed (it decreases with the increase of the collision centrality). The collective flow velocity has a more sensitive 
dependence on the collision centrality (it decreases with the decrease of the collision centrality).

\begin{figure}
\vspace*{-.45cm}
\begin{center}
\begin{tabular}{cc}

\epsfxsize=8.0cm\epsffile{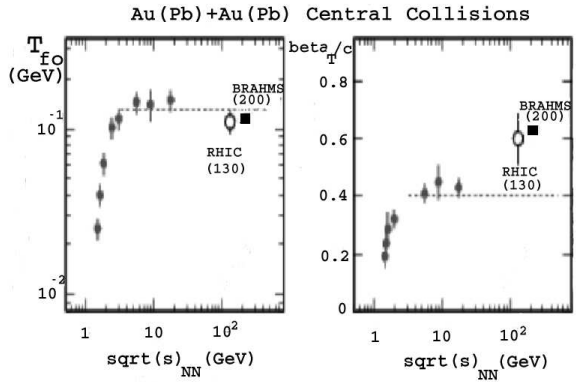}
\end{tabular}
\vspace{-0.9cm}
\end{center}
\caption[]{BRAHMS preliminary results for the most central collisions (0-10\%) in comparison with the results from other experiments at lower energies}
\label{fig7}
\end{figure}

\begin{table}
\vspace*{-12pt}
\caption[]{Freeze-out temperature and transverse collective flow velocity}
\vspace*{-14pt}
\begin{center}
\begin{tabular}{llllll}
\hline\\[-10pt]
Centrality & $T_{fo}$ & $\beta_{T}$ \\ 
\hline\\[-10pt]
$0-10\%$ & 119$\pm$1 &  0.626$\pm$0.005 \\
$10-20\%$ & 125$\pm$2 &  0.602$\pm$0.006 \\
$20-40\%$ & 127$\pm$2 &  0.583$\pm$0.008 \\
$40-60\%$ & 134$\pm$5 &  0.525$\pm$0.023 \\
\hline 
\end{tabular}
\end{center}
\end{table}

\begin{figure}
\vspace*{2.01cm}
\begin{center}
\begin{tabular}{cc}

\epsfxsize=6.0cm\epsffile{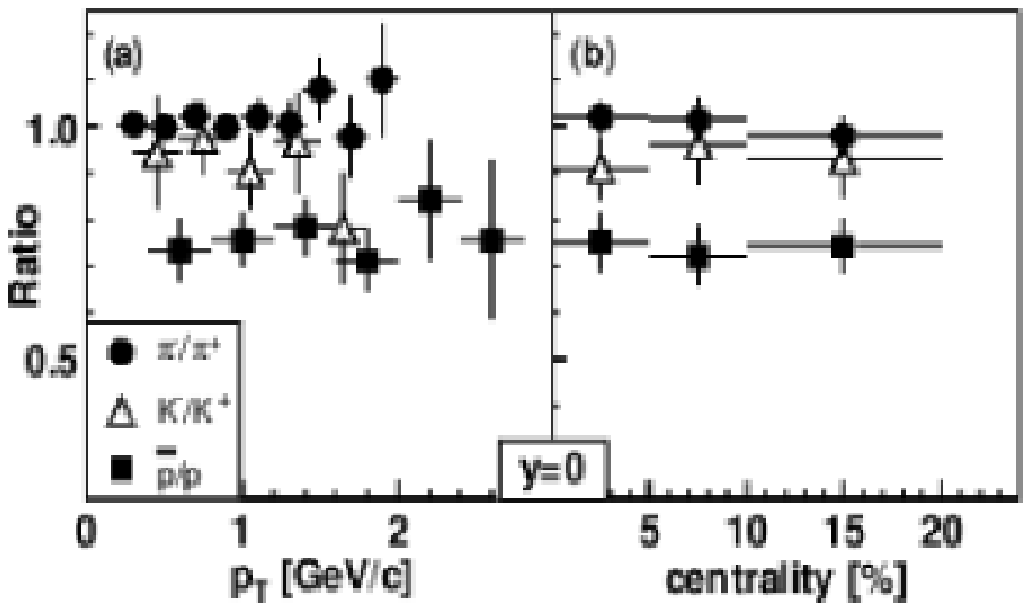}
&
\epsfxsize=6.0cm\epsffile{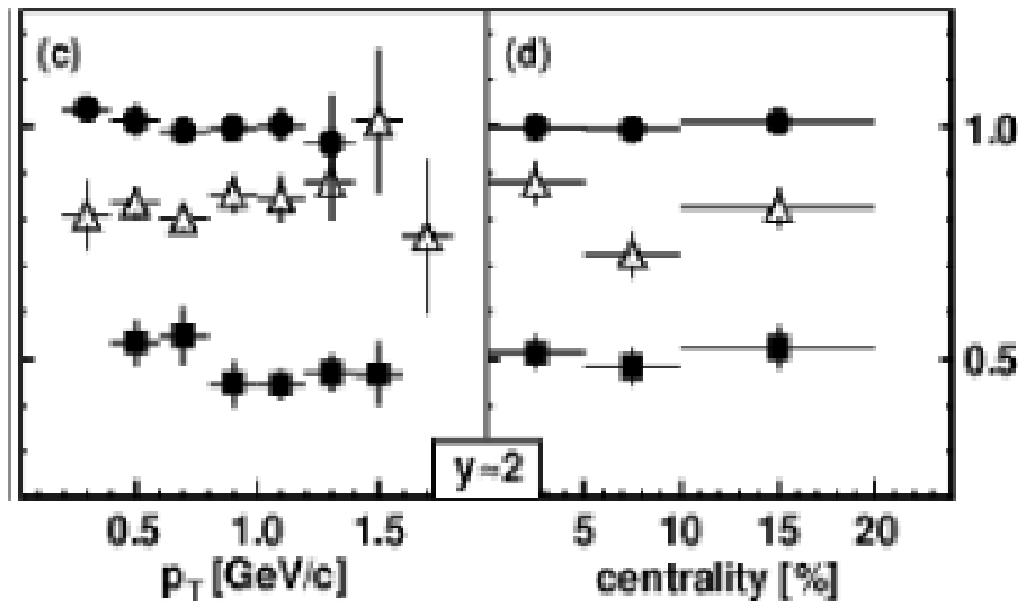} \\[0.1ex]
\end{tabular}
\vspace{-0.9cm}
\end{center}
\caption[]{Dependence of the measured antihadron to hadron ratios on transverse momentum and collision centrality}
\label{fig8}
\end{figure}

\begin{figure}
\vspace*{-.45cm}
\begin{center}
\begin{tabular}{cc}

\epsfxsize=9.0cm\epsffile{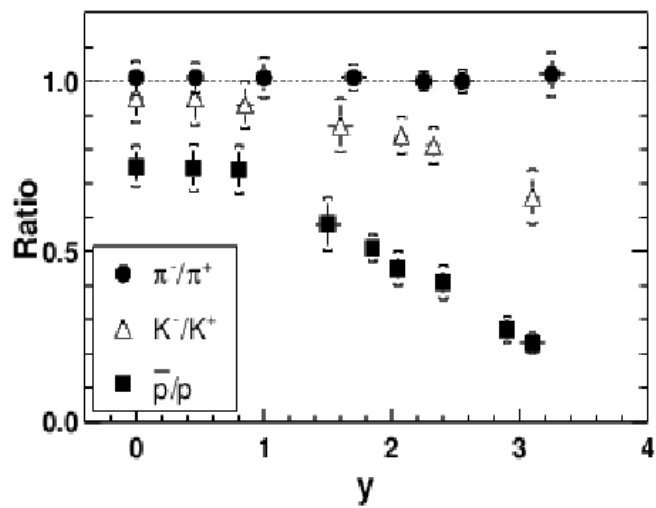}
\end{tabular}
\vspace{-0.9cm}
\end{center}
\caption[]{Antihadron to hadron ratios as a function of rapidity. Errors bars show the statistical errors while the caps indicate the combined statistical and systematic errors.}
\label{fig9}
\end{figure}

At RHIC energies the transverse collective flow velocity parameter is larger than that from collisions at lower energies 
(JINR Dubna, AGS-BNL, SPS-CERN). In the same time, the freeze-out temperature parameter is lower that obtained in 
nucleus-nucleus collisions at lower energies (see Fig.7) \cite{bib18}. 

\section{Particle ratios}

\subsection{General information}

Some information on the collision dynamics can be obtained from ratios of yields of antihadrons to hadrons, as well as from 
their dependencies on rapidity \cite{bib19}. The dependencies of the $\frac{\pi^{-}}{\pi^{+}}$, $\frac{K^{-}}{K^{+}}$ and $\frac{\bar{p}}{p}$ 
ratios on transverse momentum and collision centrality 
for Au-Au collisions at $\sqrt{s_{NN}}=200$GeV - at 2 rapidities - are presented in Fig.8.  For a given rapidity no significant dependence of the ratios 
on transverse momentum and collision centrality is observed. Therefore, an analysis of the ratios dependence on the rapidity was 
done (Fig.9). The different dynamical models do not describe such dependence. All ratios are essentially constant in the rapidity 
range $y\in[0,1]$.
The values of $K^{-}/K^{+}$ and $\bar{p}/p$ ratios obtained in Au-Au collisions at the available energies at RHIC-BNL are highest values obtained. For example, at 
y = 0.0, in Au-Au collisions at $\sqrt{s_{NN}}=200$GeV, for the most 10\% central collisions, the following average values of the three 
ratios are obtained: $<\frac{\pi^{-}}{\pi^{+}}>=1.011\pm0.006$, $<\frac{K^{-}}{K^{+}}>=0.95\pm0.05$ and $<\frac{\bar{p}}{p}>=0.75\pm0.04$.
Each ratio offers specific information. For example, the $\frac{\pi^{-}}{\pi^{+}}$ ratio can be related to the Coulomb repulsion, the 
$\frac{K^{-}}{K^{+}}$ ratio can offers information on the strangeness content, and the $\frac{\bar{p}}{p}$ ratio can says something about 
the chemical equilibration (values of different chemical potentials).

\subsection{Coulomb interaction and the pion ratio}

The Coulomb interaction between net charge of the participant region and pions can be reflected by a modification in the 
transverse momentum. The Coulomb momentum is estimated using the following relation:

\begin{equation}
p_c{\equiv}|p_{\perp}-p_{{\perp},0}|{\cong}2e^2{\frac{dN^{ch}}{dy}{\frac{1}{R_f}}}
\end{equation}
where $p_{{\perp},0}$ is the transverse momentum at freeze-out and $p_{\perp}$ is the final momentum. The freeze-out radius is:

\begin{equation}
R_f=R_{geom}+0.5{\cdot}\beta_\perp\cdot\tau_f
\end{equation}

where $R_{geom}$ is the initial (geometrical) radius of the overlap zone at the collision time and $\beta$ is the
transverse flow velocity.

Therefore, the pion ratio can be described using the following expression:

\begin{equation}
\frac{\pi^-}{\pi^+}=\left\langle\frac{\pi^-}{\pi^+}\right\rangle\frac{p_{\perp}+p_c}{p_{\perp}-p_c}\exp{\left(\frac{m_{\perp}^--m_{\perp}^+}{T}\right)}
\end{equation}
where $m_\perp^\pm=\sqrt{m^2+\left(p_{\perp}{\pm}p_c\right)^2}$.

\begin{figure}
\vspace*{-.45cm}
\begin{center}
\begin{tabular}{cc}

\epsfxsize=8.0cm\epsffile{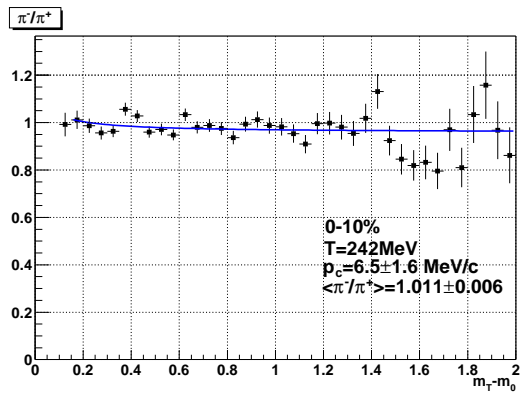}
\end{tabular}
\vspace{-0.9cm}
\end{center}
\caption[]{The pion ratio dependence on the transverse mass for the most central Au-Au collisions at 200 AGeV (0-10\%) - {\it {preliminary data}}}
\label{fig10}
\end{figure}

The Coulomb effects in pion spectra are sensitive to the degree of stopping, the distribution of the positive charge and flow
velocity of the participant region. The experimental results shown in Fig.10 obtained in Au-Au collisions at reflect a reduced
Coulomb effect ($p_{c}=6.5\pm1.9MeV/c$). This behaviour can be related to the higher flow velocities of the nuclear matter for these collisions, in
comparison with similar collisions at lower energies ($p_c^{AGS}\cong20MeV/c$, $p_c^{SPS}\cong10MeV/c$).

\subsection{Thermal interpretation of the kaon ratio dependence on the antiproton-proton ratio}

The baryon chemical potential can be related to the ratio $\frac{\bar{p}}{p}$ by the following relation:

\begin{equation}
\frac{\bar{p}}{p}=exp(-\frac{2\mu_{B}}{T})
\end{equation}

\begin{figure}
\vspace*{-.45cm}
\begin{center}
\begin{tabular}{cc}

\epsfxsize=7.0cm\epsffile{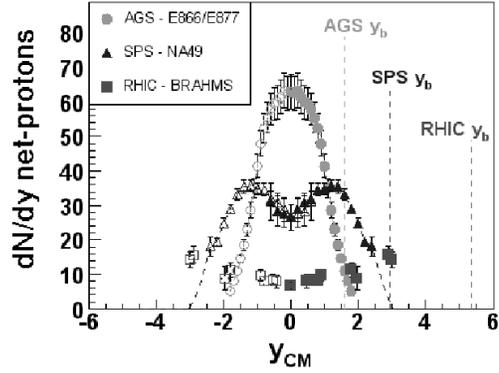}
\end{tabular}
\vspace{-0.9cm}
\end{center}
\caption[]{Net-proton rapidity densities for central heavy ion collisions at AGS, SPS and RHIC}
\label{fig11}
\end{figure}

Using the experimental values of the ratio obtained in the BRAHMS experiment ($0.64\pm0.04(stat)\pm0.06(syst)$ 
for Au-Au collisions at $\sqrt{s_{NN}}=130GeV$, $0.75\pm0.03(stat)\pm0.05(syst)$ for the same collisions at 
$\sqrt{s_{NN}}=200GeV$, respectively), as well as the following dependence of the baryon chemical 
potential on rapidity:

\begin{equation}
\frac{\mu_{B}}{T}=a{(1+y)^b}
\end{equation}

with $a=0.09\pm0.01$ and $b=1.69\pm0.21$, a increase of the baryon chemical potential with the increase 
of the rapidity has been observed (Table III). The much reduced net-baryon density observed at 
mid-rapidity ($10\pm2$) in comparison with forward rapidities (see Fig.11) could explain this behaviour. 
Very different collision evolutions at different energies can be observed, also.  

\begin{table}[hb]
\vspace*{-12pt}
\caption[]{The dependence of the baryon chemical potential on the rapidity in Au-Au collisions at 200GeV}
\vspace*{-14pt}
\begin{center}
\begin{tabular}{llllll}
\hline\\[-10pt]
y & $\mu_{B}[MeV]$  \\ 
\hline\\[-10pt]
0 & 28.8$\pm$2.1 \\
1.5 & 46.7$\pm$8.9 \\
2.1 & 66.8$\pm$9.4 \\
3.1 & 123.1$\pm$10.9 \\
\hline 
\end{tabular}
\end{center}
\end{table}

For the $\frac{K^{-}}{K^{+}}$ ratio a similar dependence can be written:

\begin{equation}
\frac{K^{+}}{K^{-}}=e^{\frac{2\mu_{B}}{3T}-\frac{2\mu_{s}}{T}}={(\frac{p}{\bar{p}})}^{1/3}e^{-\frac{2\mu_{s}}{T}}
\end{equation}

\begin{figure}
\vspace*{-.45cm}
\begin{center}
\begin{tabular}{cc}

\epsfxsize=7.0cm\epsffile{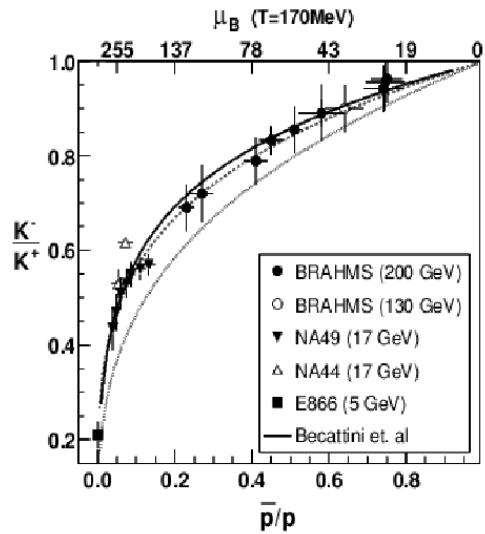}
\end{tabular}
\vspace{-0.9cm}
\end{center}
\caption[]{Correlation between kaon and baryon ratios (both systematic and statistic errors are included). Statistical model prediction of Becatini et al is shown by continuous line}
\label{fig12}
\end{figure}

In Fig.12 the dependence:

\begin{equation}
\frac{K^{-}}{K^{+}}=(\frac{\bar{p}}{p})^a
\end{equation}

is shown. For SPS experimental results the following results are obtained: $a=0.20\pm0.01$, and for 
BRAHMS experimental results $a=0.24\pm0.02$. Both are obtained assuming T =170 MeV. This choice is 
in agreement with the results obtained supposing the following energy dependence of the chemical 
temperature:

\begin{equation}
T_{ch}=a-be^{-c\sqrt{s}}
\end{equation}

with $a=172.3\pm2.8$, $b=149.5\pm5.7$, $c=0.20\pm0.03$. For Au-Au collisions at $\sqrt{s_{NN}}=130GeV$ 
and $\sqrt{s_{NN}}=200GeV$ a common value is obtained, namely: $T_{ch}=172.3\pm2.8MeV$. This results, 
together with experimental results obtained at lower energies (see Fig.13), suggest that the 
chemical temperature saturates close to the critical temperature of 170 MeV extracted from lattice 
QCD calculations. 

\begin{figure}
\vspace*{-.45cm}
\begin{center}
\begin{tabular}{cc}

\epsfxsize=8.0cm\epsffile{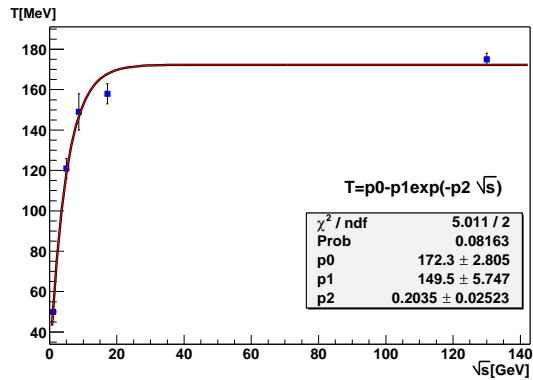}
\end{tabular}
\vspace{-0.9cm}
\end{center}
\caption[]{The energy dependence of the chemical temperature}
\label{fig13}
\end{figure}

It is important to stress here that the values of the baryon chemical potential obtained in Au-Au
collisions at the two energies ($43.5\pm3.2MeV$, $28.8\pm2.1$ respectively) are in agreement with 
the following dependence \cite{bib21}:

\begin{equation}
\mu_{B}(s)=\frac{a}{1+\frac{\sqrt{s}}{b}}
\end{equation}

with $a=0.967\pm0.032GeV$, $b=6.138\pm0.399GeV$.

As general remarks on the dynamic information obtained from the behaviours of the antihadron to hadron 
ratios it is possible to affirm that the experimental values obtained in Au-Au collisions are the 
highest reported. The increase in $\frac{K^{-}}{K^{+}}$ and $\frac{\bar{p}}{p}$ ratios can be consistent 
with the increase in collision transparency in comparison with lower energies. All ratios are constant, 
in the limit of the experimental errors, in the rapidity range $y\in[0,1]$. This behaviour is in agreement 
with the large plateau observed in the charged particle rapidity distributions. The ratios permit 
to estimate baryon chemical potential and a chemical freeze-out temperature. At the RHIC top energy 
the value of the baryon chemical potential is lowest obtained up to now. At the same energy the baryon 
chemical potential increase significantly from mid-rapidity to forward rapidities. The net-baryon 
content is lowest at the mid-rapidity and could explain the behaviour of the baryon chemical potential. 
The correlation between $\frac{K^{-}}{K^{+}}$ and $\frac{\bar{p}}{p}$ ratios is well described by a 
statistical model.

\section{Some concluding remarks}

With the BRAHMS experiment from RHIC-BNL has been obtained a wealth of new and detailed 
information on relativistic heavy ion collision dynamics. These experimental data and results offer 
a tool to explore and quantify very hot and dense nuclear matter. The net-baryon density is very small 
(around 10), and the corresponding chemical potential is also small (around 29 MeV).

The system exhibits a large transverse and longitudinal expansion with the azimuthally asymmetries 
being large, reflecting the initial partonic distributions. This behaviour is consistent with the 
assumption that the system has reached a hydrodynamic limit. This limit can be used to explore the
equation of state of the hot and dense nuclear matter.  Suppression of the high transverse momentum 
particles relative to elementary proton-proton collisions was observed in central Au-Au collisions. 

The heavy ion data from RHIC are consistent with formation of a hot dense system that exhibits 
hydrodynamic behaviour with rapid transverse and longitudinal expansion.  Also, the system absorbs 
high transverse momentum probes corresponding to a large gluon density in the initially formed 
system. Taking into a account the value of the net-baryon density it is possible to affirm that is 
an almost baryon free system.

\section*{Acknowledgements}

This work was supported by the division of Nuclear Physics of the Office of Science of the U.S. DOE, 
the Danish Natural Science Research Council, the Research Council of Norway, the Polish State Committee 
for Scientific Research and the Romanian Ministry of Education and Research.

One of the authors (Alexandru Jipa) wishes to thank to Dr. Tamas Csorgo and Dr. Peter Levai for 
invitation to present a lecture at the Third Winter School of Relativistic Heavy Ion Collisions, 
Budapest, Hungary, as well as to the members of the BRAHMS Collaboration for their support.

\section*{Notes}  
\begin{notes}
\item[a]
Permanent address: Faculty of Physics, University of Bucharest, \\
P.O.Box MG-11, RO 077125 Bucuresti-Magurele, ROMANIA\\ 
E-mail: jipa@brahms.fizica.unibuc.ro
\end{notes}

\vfill\eject

\begin{thebibliography}{99}  
\bibitem{bib1}D.Beavis et al - BRAHMS Conceptual Design Report, BNL-62018, October 1994.

\bibitem{bib2}Fl.Videbaek, Talk at Workshop UIC, June 14-17, 1998, World Scientific, Singapore, 1999.

\bibitem{bib3}F.Videbaek for the BRAHMS Collaboration, International Conference "Quark Matter 2001",
Stony Brook and Brookhaven, January 2001, USA,  published in {\it Nuclear Physics}  {\bf A698} (2002) 29c-38c.

\bibitem{bib4}Al.Jipa for the BRAHMS Collaboration, {\it Rom.Rep.Phys.} {\bf 53(3-8)} (2001) 323-334.

\bibitem{bib5}M.Adamczyk et al (BRAHMS Collaboration), {\it Nucl.Inst.Meth.Phys.Res.} {\bf A499} (2003) 437-468.

\bibitem{bib6}C.Adler et al,{\it Nucl.Inst.Meth.Phys.Res.} {\bf A499} (2003).

\bibitem{bib7}I.G.Bearden et al (BRAHMS Collaboration), {\it Phys.Rev.Lett.} {\bf 87} (2001) 112305.
 
\bibitem{bib8}I.G.Bearden et al (BRAHMS Collaboration), {\it Phys.Rev.Lett.} {\bf 88} (2002) 202301.

\bibitem{bib9}M.Gyulassy, Xian-Nian Wang, Preprint LBL, LBL-34246(2000).

\bibitem{bib10}Al.Jipa, Analysis Note 17, BRAHMS Analysis Notes (www.bnl.gov/brahms/).

\bibitem{bib11}B.Back ey al (PHOBOS Collaboration), {\it Phys.Rev.Lett.} {\bf 85} (200) 3100.

\bibitem{bib12}J.D.Bjorken, {\it Phys.Rev.D} {\bf 27} (1983) 140.

\bibitem{bib13}Al.Jipa, {\it J.Phys.G:Nucl.Part.Phys.} {\bf 22} (1996) 231; Nu Xu for N44 Coll., {\it Nucl.Phys.A} {\bf610} (1996) 175c

\bibitem{bib14}Al.Jipa for the BRAHMS Collaboration, {\it Rom.Rep.Phys.} {\bf 53(3-8)} (2001) 367-379.

\bibitem{bib15}I.G.Bearden et al, {\it Phys.Rev.Lett.} {\bf 78} (1997) 2080.

\bibitem{bib16}J.Rayford Nix,{\it Phys.Rev.C} {\bf 58} (1998) 2303.

\bibitem{bib17}T.Ullrich, Preprint nucl-ex/0211004.

\bibitem{bib18}N.Xu, M.Kaneta,{\it Nucl.Phys.} {\bf A698} (2002) 306c.

\bibitem{bib19}I.G.Bearden et al (BRAHMS Collaboration){\it Phys.Rev.Lett.} {\bf 90} (2003) 102301.

\bibitem{bib20}F.Becatini et al {\it Phys.Rev.C} {\bf 64} (2001) 024901

\bibitem{bib21}P.Braun-Munzingen, K.Redlich, J.Stachel, Preprint nucl-th/0304013

\end{thebibliography}
\end{document}